\documentclass[a4paper,11pt]{article}

\usepackage{url}
\usepackage{amssymb,amsmath,amsthm}
\usepackage{natbib}
\usepackage{graphicx}
\usepackage{transparent}
\usepackage[font={singlespacing,small,it}]{caption}
\usepackage{hyperref}
\textfloatsep = \floatsep
\intextsep = \floatsep
\newcommand{\beq}{\begin{equation}}
\newcommand{\eeq}{\end{equation}}

\begin{document}
\font\myfont=cmr12 at 15pt
\title{{\myfont Energy Disaggregation for SMEs using\\ Recurrence Quantification Analysis}}
\date{February, 2018}
\author{Laura Hattam and Danica Vukadinovi\'c Greetham\footnote{Department of Mathematics and Statistics, University of Reading, Whiteknights, Reading, UK}}
\maketitle
\title{}

\begin{abstract}
Energy disaggregation determines the energy consumption of individual appliances from the total demand signal, which is recorded using a single monitoring device. There are varied approaches to this problem, which are applied to different settings. Here, we focus on small and medium enterprises (SMEs) and explore useful applications for energy disaggregation from the perspective of SMEs. More precisely, we use recurrence quantification analysis (RQA) of the aggregate and the individual device signals to create a two-dimensional map, which is an outlined region in a reduced information space that corresponds to `normal' energy demand. Then, this map is used to monitor and control future energy consumption within the example business so to improve their energy efficiency practices. In particular, our proposed method is shown to detect when an appliance may be faulty and if an unexpected, additional device is in use.

\end{abstract}
\section{Introduction}
In recent years, several active energy management solutions have been developed that are targeted at large manufacturers with integrated systems. Also, a wide range of household micro-systems are becoming available, which take advantage of the current smart meter roll-out. In contrast, very few meso-level systems, i.e. those targeted at small and medium enterprises (SMEs) are presently available.  A recent research by the UK government\footnote{\url{https://www.gov.uk/government/publications/smart-metering-in-
non-domestic-premises-early-research-findings}} has  shown that a very small number of SMEs are actively engaged in energy management, including the use of smart or advanced meter data. During the last twenty years, the main consistent barriers are the lack of funds for capital equipment and the lack of in-house expertise to monitor and manage systems. However, it is clear that, in principle, SMEs could achieve substantial energy cost savings simply through some changes in their usage behaviour.

Responding to this gap in the market, here we will outline a method for monitoring and controlling energy usage within a small to medium sized business. The energy consumption of an example dry cleaners over a six week period is used here to demonstrate the proposed technique. From current data collected on site, a disaggregation algorithm is devised so that certain electrical devices and/or device combinations can be identified from the aggregate energy demand. This allows for the behaviour of select devices to be monitored within an example environment and as a result, form control solutions to improve the energy efficiency of its operations.

Since individual device monitoring is expensive, energy disaggregation is an extremely useful tool for energy management as it enables the monitoring of multiple devices with only one smart meter. The main technique adopted here to perform energy disaggregation is recurrence quantification analysis (RQA) of a recurrence plot, which was firstly proposed by \citet{abi92}. More specifically, a recurrence plot is constructed to reveal the repeating patterns within a complex time series that are initially somewhat hidden. Then, RQA is performed to quantify the recurrent structures illustrated by these plots, where time dependent RQA variables are computed.

The chosen approach is motivated by the findings of \citet{fab05} and \citet{mar13}. They demonstrated that by computing RQA variables for some time series, significant events and phase transitions could be detected, such as financial bubbles, financial crashes and transitions in climate systems. In addition, \citet{mos16} applied RQA to various time series that were associated with different flow patterns. Consequently, certain RQA measurements were selected that best captured the behaviour of each flow. Next, principal component analysis (PCA) reduced the number of components down to two. Lastly, as a function of these two components, individual, approximate zones were outlined that corresponded to the different flow patterns. We build upon the studies of \citet{mos16} here by using a similar approach for the new application of energy disaggregation. Our proposed method works for lower data resolution, and also uses PCA to map selected RQA measurements to two components. Then, distinct regions within the reduced space are highlighted to represent different usage behaviours, which allows for an intuitive and easy visualisation of the data in the form of 2-d maps. This is again very relevant to SMEs as it enables the engagement of non-expert users, whilst ensuring that the equipment is lower in cost and the data storage requirements are minimised.

The outline of this paper is the following: in Section \ref{pw} we give a short overview of relevant literature on SME energy management, other energy disaggregation methods, and direct the reader to RQA methodology papers and its application in various fields; then in Section \ref{dcdat} we describe the data used for our case study;  in Section \ref{meth}, the details of our method are given, and in Section \ref{map}, the visualisation process is described; the method results obtained for the case-study data are discussed in Section \ref{res}; finally, we conclude in Section \ref{con}, where details of future work are given.

\section{Previous work\label{pw}}
With increasing interest in energy efficiency measures, most of the attention has naturally been focused on households and on the largest commercial energy consumers, which are on opposite ends of the spectrum in terms of energy consumption (see \citet{sch16}). Conversely, enterprises sitting somewhere in between these two groups have been somewhat overlooked. Due to the heterogeneous nature of SMEs, it is difficult to develop a generalised approach, where the existing literature is split over varying industrial sectors. In addition, trying to copy the solutions from larger industries does not always work in the context of SMEs. For example, \citet{kannan03} examined a case study on SME's energy management practices, which revealed that SMEs are unable to appoint a dedicated energy manager since the potential energy savings do not outweigh the position costs. \citet{tri12} investigated different barriers to energy efficiency for SMEs and identified that the major three are access to capital, lack or imperfect information on cost-efficient energy interventions and the form of information. They also warned against bundling up together small, medium and large sized enterprises as they usually have different issues. \citet{bla14} studied energy efficiency measures among SMEs in the USA, where top operations managers are identified as being key to enforcing new energy saving measures. From all of this, it is safe to assume that the personnel who will be adopting the new practices is in general a non-expert, and therefore, any methods developed for SMEs need to be accessible and affordable in order to be considered.

Energy disaggregation is an active field of research due to, amongst other reasons, its relevance to energy efficiency strategies. \citet{car13} argued that numerous user-benefits resulted by making available the appliance-level energy data, and the combination of algorithms and smart meters is the most cost-effective/scalable solution to obtaining this data. It was also claimed that disaggregation is often used to provide recommendations (e.g. audits or appliance replacements), to detect malfunctions, and to enable new behaviours. Again, on a commercial level, these are mostly untapped savings in SMEs.

Non-intrusive appliance load monitoring (NALM, sometimes NILM or NIALM, see \citet{zoh12}) is the disaggregation of a household's total energy consumption into individual device consumption, which uses analytical techniques only. There is currently a lot of research in machine learning methods for NALM (for an example, refer to \citet{tab17}). However, these algorithms that disaggregate power loads at low sampling rates of several minutes or half hourly are still not accurate enough, nor practical, as they require substantial customer input and long training periods. Two low-rate NALM techniques, one combining  k-means clustering and Support Vector Machine, and another requiring a database of appliance signatures created using publicly available datasets are outlined by \citet{alt16}. Due to the heterogeneity of SMEs, it is expected that the creation of a signature database that encompasses all the different appliance scenarios will be challenging and time consuming, when compared to household appliances that are widely used and mass produced.

We are contributing to this field of research by introducing RQA to SME energy disaggregation, which to our knowledge is a novel application for RQA. We will show that this method does not require any machine learning, it works in low resolutions and the results can be easily visualised. These features allow us to create an intuitive non-expert interface for monitoring/controlling energy usage, which will be especially useful in a SME context.

Recurrence is a fundamental property of dynamical systems, and captures the system's behaviour in phase space. In order to facilitate visualisation and analysis of recurrences within a time series, \citet{eck87} introduced recurrence plots. Later, the quantification of recurrence plots and RQA was proposed in the early $1990$s, and has been shown to be extremely useful for detecting transitions in the system dynamics, as well as other relevant system properties. For the mathematical details of RQA and an overview of its applications across economy, physiology, neuroscience, earth sciences, astrophysics and engineering, refer to \citet{mar07}.

\section{Dry Cleaners Data\label{dcdat}}
\begin{figure*}
\centering
\includegraphics[width=3.3in,height=2.7in]{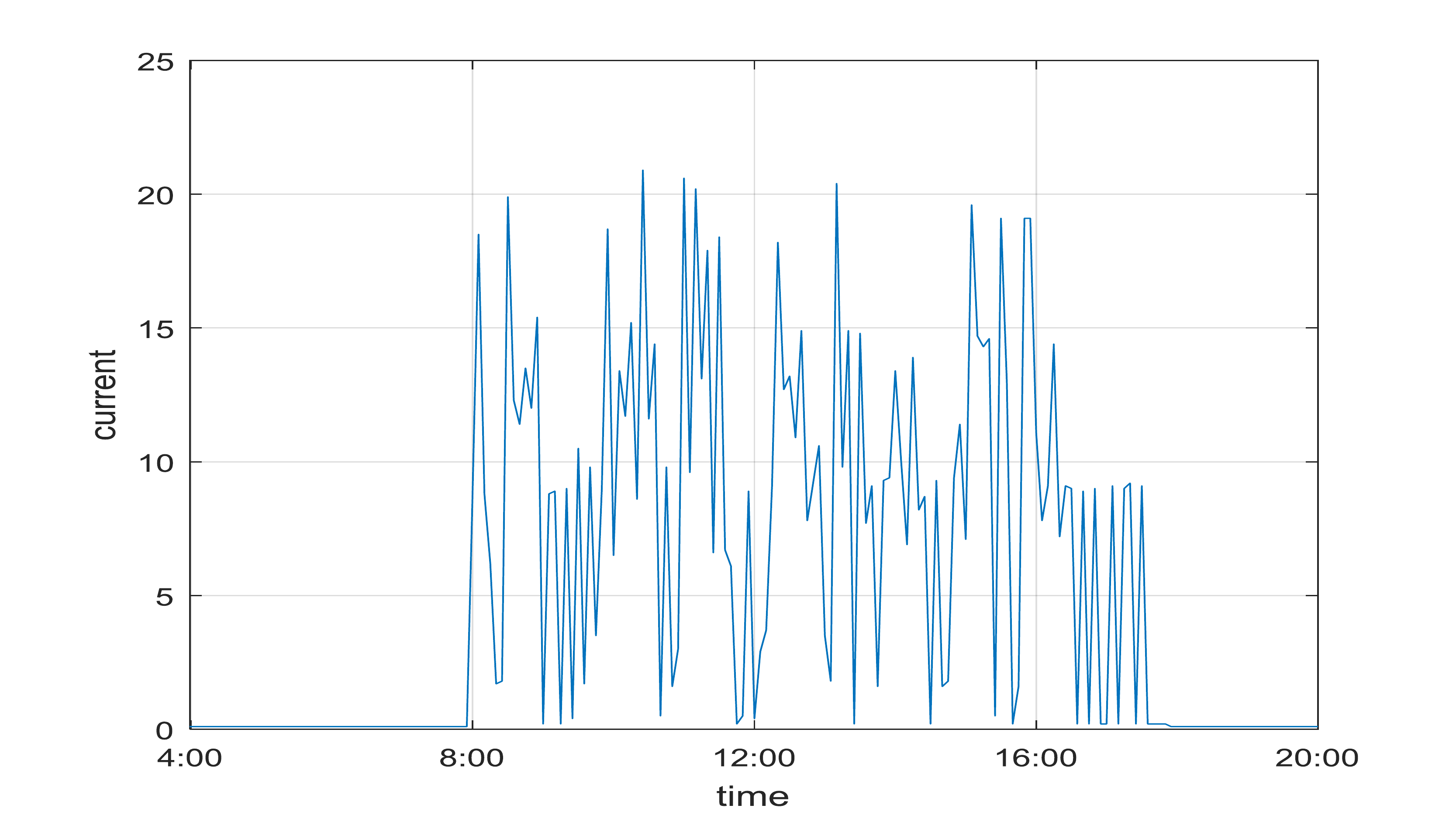}\includegraphics[width=3.3in,height=2.7in]{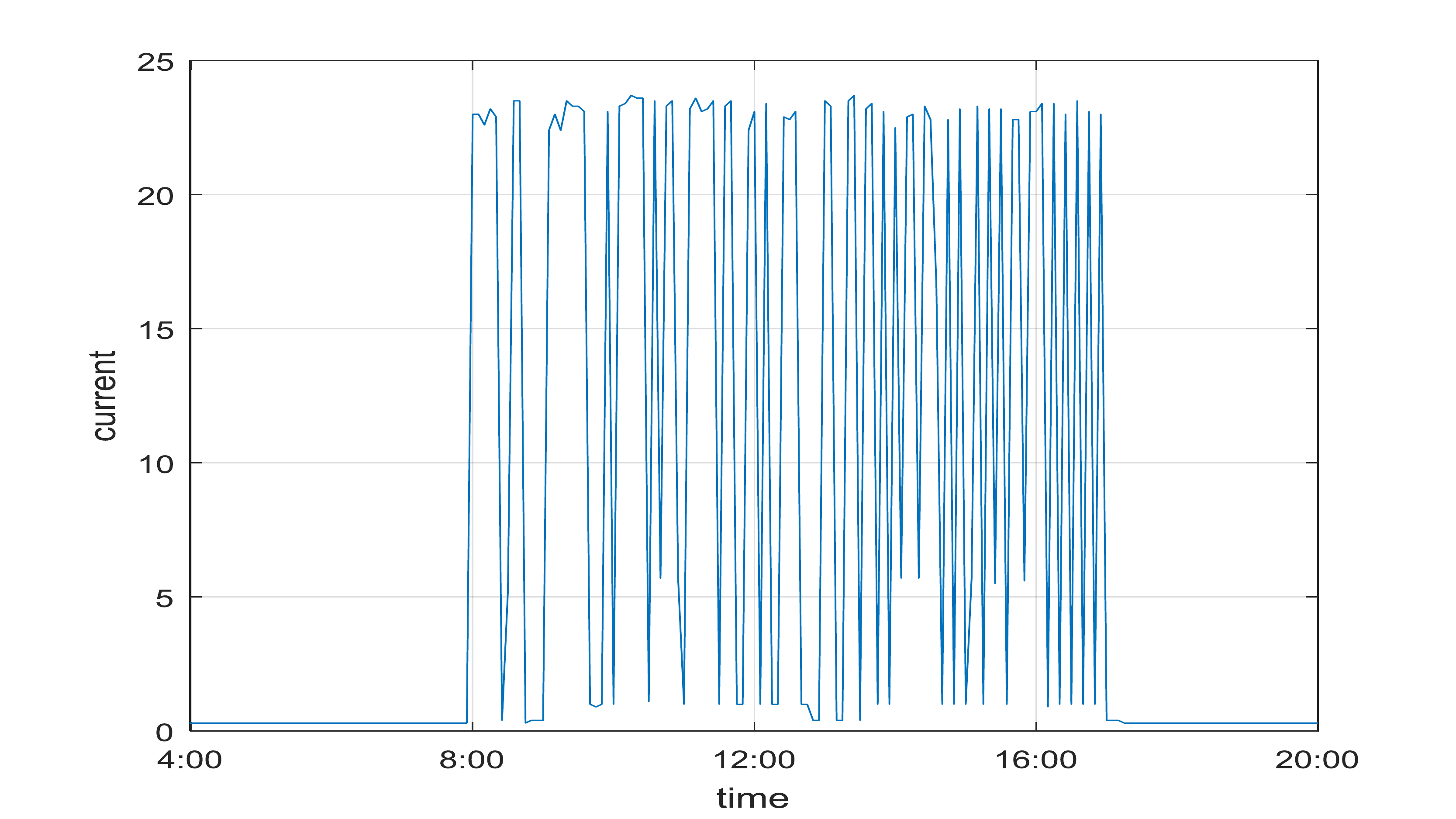}\\
\includegraphics[width=3.3in,height=2.7in]{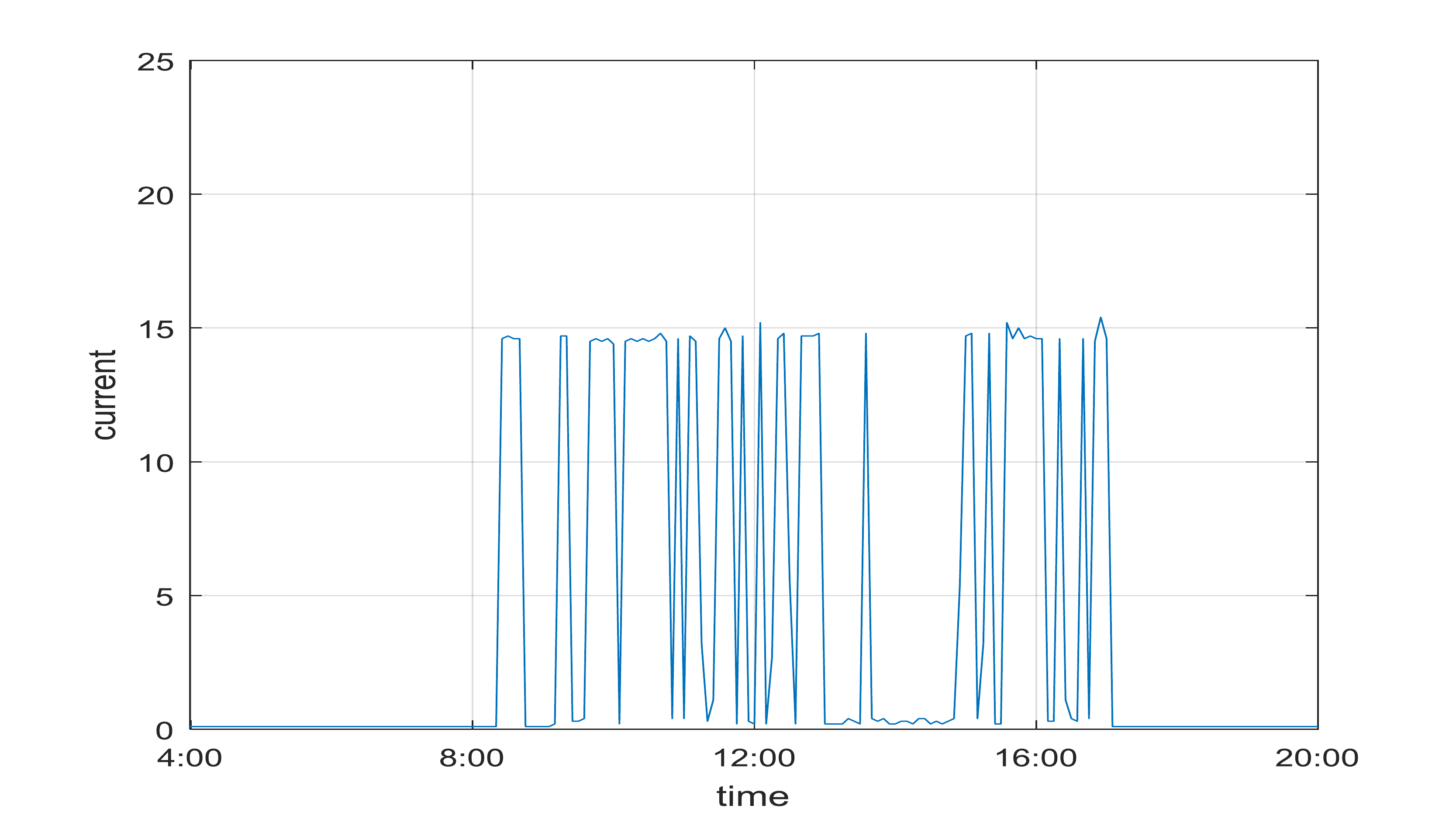}\includegraphics[width=3.3in,height=2.7in]{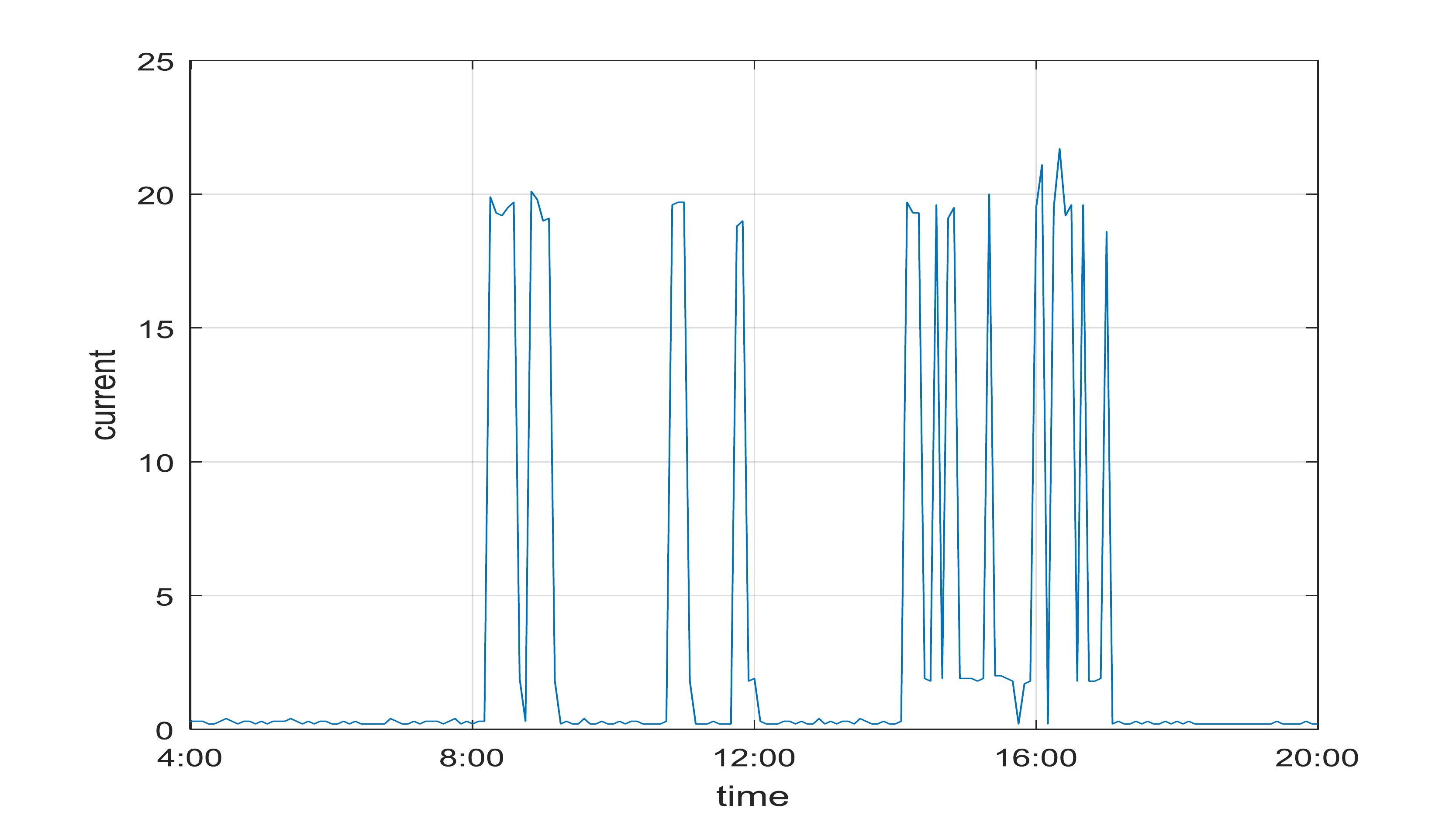}
\caption{Example daily current demand (Amps) for the devices: cleaner (top left), iron (top right), table press (bottom left), dryer (bottom right).\label{curr_dev} }
\end{figure*}

For six weeks, energy data was recorded at the dry cleaners, with current readings every five minutes, for the following appliances:
\begin{enumerate}
  \item Dryer- single phase
  \item Table press- three phase
  \item Iron- three phase
  \item Cleaner- three phase
  \item Aggregate- three phase
\end{enumerate}
We choose to focus now and for the remainder of the paper on one phase which has all four devices connected. Examples of the daily current demands for this phase and the various appliances are exhibited in Figure \ref{curr_dev}. By analysing this data over the monitored six weeks, it is found that the iron and cleaner switch on and off continuously throughout the day, every working day. In contrast, the dryer switches on every working day, except sometimes it is off for parts of the day. Lastly, the table press is used every day, switching on and off continuously throughout the day, except for nine non-consecutive days when it remains switched off all day.

Note that due to the energy information being defined at varying times, with readings every five minutes i.e. device $A$ and $B$ begin recordings at, say, $00$:$01$ and $00$:$03$ respectively, we linearly interpolate the data so that it is then defined every minute. As well, the dry cleaners is closed on Sundays, and therefore, the data associated with these closure days is omitted.

\section{Energy Disaggregation Method\label{meth}}

For a data set consisting of the current profiles for the individual appliances and the aggregate demand, we apply the following energy disaggregation method:
\begin{enumerate}
  \item The recurrence plots are constructed for the aggregate signal and the individual device signals.
  \item Next, RQA variables are computed using a sliding window along the recurrence plot diagonal.
  \item The RQA measurements found in step $2$ are then associated with the particular appliance or appliance combination when `on' (RQA data corresponding to a device's `off' state is omitted from the analysis).
  \item This process is repeated for a number of weeks so that the range of RQA values for each device/device combination is obtained.
  \item $N$ RQA variables are chosen that best characterise the various appliance states, where $N>2$ is some integer.
  \item PCA is next performed to reduce this problem from $N$ components to two.
  \item The RQA information for each appliance/appliance combination is plotted as a function of the two components determined in step $6$.
  \item Then, distinct regions are highlighted in this two dimensional space to correspond to each device/device combination. Furthermore, the union of these outlined regions creates a map to represent a zone of `normal' energy usage, which can be later used for control solutions.
\end{enumerate}
The theory associated with this method is further detailed in Sections \ref{rp}-\ref{pca}. Note that when constructing a recurrence plot for the one-dimensional time series $\textbf{Y}=(Y_1,Y_2,\ldots,Y_n)$ of length $n$, often $\textbf{Y}$ is initially embedded in a higher dimensional space (see \citet{kan03}). However, through experimentation with the dry cleaners data, it was concluded that embedding is unnecessary here.
\subsection{Recurrence Plots\label{rp}}
To create the recurrence plot for the time series $\textbf{Y}=(Y_1,Y_2,\ldots,Y_n)$, firstly the distance matrix (DM) is calculated, which is a matrix of points $DM(i,j)$, where $i,j=1,2,\ldots,n$. The value at entry $DM(i,j)$ is defined as the distance between $Y_i$ and $Y_j$ i.e. $DM(i,j)=|Y_i-Y_j|$, where the matrix aligns vertically with time step $i$ and horizontally with time step $j$. This means that the elements along the diagonal are equal to zero and the matrix is symmetric about the diagonal. From the DM, the recurrence plot is determined, which is a matrix of points $R(i,j)$ defined as
\beq
R(i,j)=H(\epsilon-DM(i,j)),
\eeq
where
\beq
H(x)=\begin{cases}
    0, & \text{if $x<0$},\\
    1, & \text{otherwise}.
  \end{cases}
  \eeq
Thus, the entry $R(i,j)$ equals one when the distance between $Y_i$ and $Y_j$ is within the threshold $\epsilon$. When this occurs, $(i,j)$ is considered recurrent and is labelled as a recurrence point. Now, since $Y_i$ represents some current level and if $R(i,j)=1$, then the point $Y_j$ is within some neighbouring region of $Y_i$ and therefore, the current level at time step $i$ is similar to the current level at time step $j$. Therefore, using recurrence plots allows us to uncover any reoccurring behaviours within our current time series.
\subsection{Recurrence Quantification Analysis\label{rqa}}
RQA is the technique applied to quantify the repeating patterns observed within the recurrence plots. The following variables are the measures considered here:
\beq REC=100\times\frac{(\text{no. of recurrence points within the window})}{W(W-1)},\eeq
which is the percentage of all points within the square window of size $W$ that are recurrent.
\beq
         DET=100\times \frac{(\text{no. of recurrence points forming diagonal line segments})}{(\text{no. of recurrence points})},\eeq
 which is the percentage of the recurrence points that form line segments parallel to the matrix diagonal. These structures relate to repeating or deterministic patterns within the system.
\beq
   ENT=-\sum P \log_2(P),\eeq
where $P$ is the probability distribution for the length of the diagonal line segments. $ENT$ is the Shannon entropy of the distribution of diagonal line segments. This variable represents complexity, where small and large $ENT$ signify periodic and unpredictable behaviour respectively. 
\beq
LAM=100\times\frac{(\text{no. of recurrence points forming vertical line segments})}{(\text{no. of recurrence points})},
\eeq
which is the percentage of the recurrence points that form vertical line segments.  These structures represent stationary behaviour.
\beq
TT=\text{average length of the vertical line segments},\eeq
which is the mean length of the vertical line segments.

Since our data consists of a set of long time series, each series is separated into a number of equally sized subseries referred to as epochs. Then, the RQA variables are calculated at every epoch. As a result, we observe how the RQA measurements vary as a function of time. This process is achieved here by sliding a square window of size $W$ along the diagonal of the recurrence plot, whilst performing RQA at every time increment.

\subsection{Principal Component Analysis\label{pca}}
For information expressed in terms of many variables, PCA calculates an orthogonal basis to describe this data so that the number of variables can be minimised. This technique is applied here to reduce the number of RQA variables representing each device/device combination.

To begin, we form a set of $n$ vectors $U_i$ of length $m$, where $i=1,2,\ldots,5$ relates to each device type (cleaner, iron, table press, dryer, aggregate), $n=5$ is the number of device states and $m=5$ is the number of RQA variables considered. The vector $U_i$ is defined as
\beq
U_i=mean(RQA_i),\eeq
where
\beq
RQA_i=\left[REC_i,DET_i,ENT_i,LAM_i,TT_i\right],\eeq
which is a matrix that corresponds to the RQA information computed for device $i$ using the method of a sliding window, as detailed in Section \ref{rqa}.

Next, the average feature of the set $U_i$ is found, which is given by
\beq
\psi=\frac{1}{n}\sum_{i=1}^{n}U_i,
\eeq
and the covariance matrix is obtained using
\beq
C=\frac{1}{n}\sum_{i=1}^n \phi_i\phi_i^T=AA^T,
\eeq
where $\phi_i=U_i-\psi$ and $A=[\phi_1\;\phi_2 \cdots \phi_n]$.

Let us now refer to the eigenvectors of $C$ as $V_j$. These vectors form the orthogonal basis for our information space. If their corresponding eigenvalues are small, then these eigenvectors are omitted and therefore, the size of the data space is decreased. The remaining eigenvectors next form the matrix $E=[V_1\; V_2 \cdots V_r]$, where $r$ is the number of eigenvectors. Now, the RQA data associated with device $i$, say the vector $RQA_{i,j}$ of length $m$, can be projected onto the reduced space such that
\beq
w_{i}^j=(RQA_{i,j}-\psi)^T E,\label{wij}\eeq
which is a vector of length $r<m$. Here, we ensure that $r=2$ so that the RQA device information is projected onto a two dimensional space with the coordinates defined by (\ref{wij}). Consequently, a 2-d map can then be outlined to depict typical energy demand at the dry cleaners.

Note that the optimal choice for $\epsilon$ and $W$ (refer to Sections \ref{rp}-\ref{rqa}) corresponds to when the individual device clusters become mostly distinct/separate within this reduced space. Therefore, the selection of the parameters $\epsilon$ and $W$ is determined through experimentation and visual assessment of the clusters.

\section{The 2-D Map\label{map}}
\begin{figure*}
\centering
\includegraphics[width=3.3in,height=2.7in]{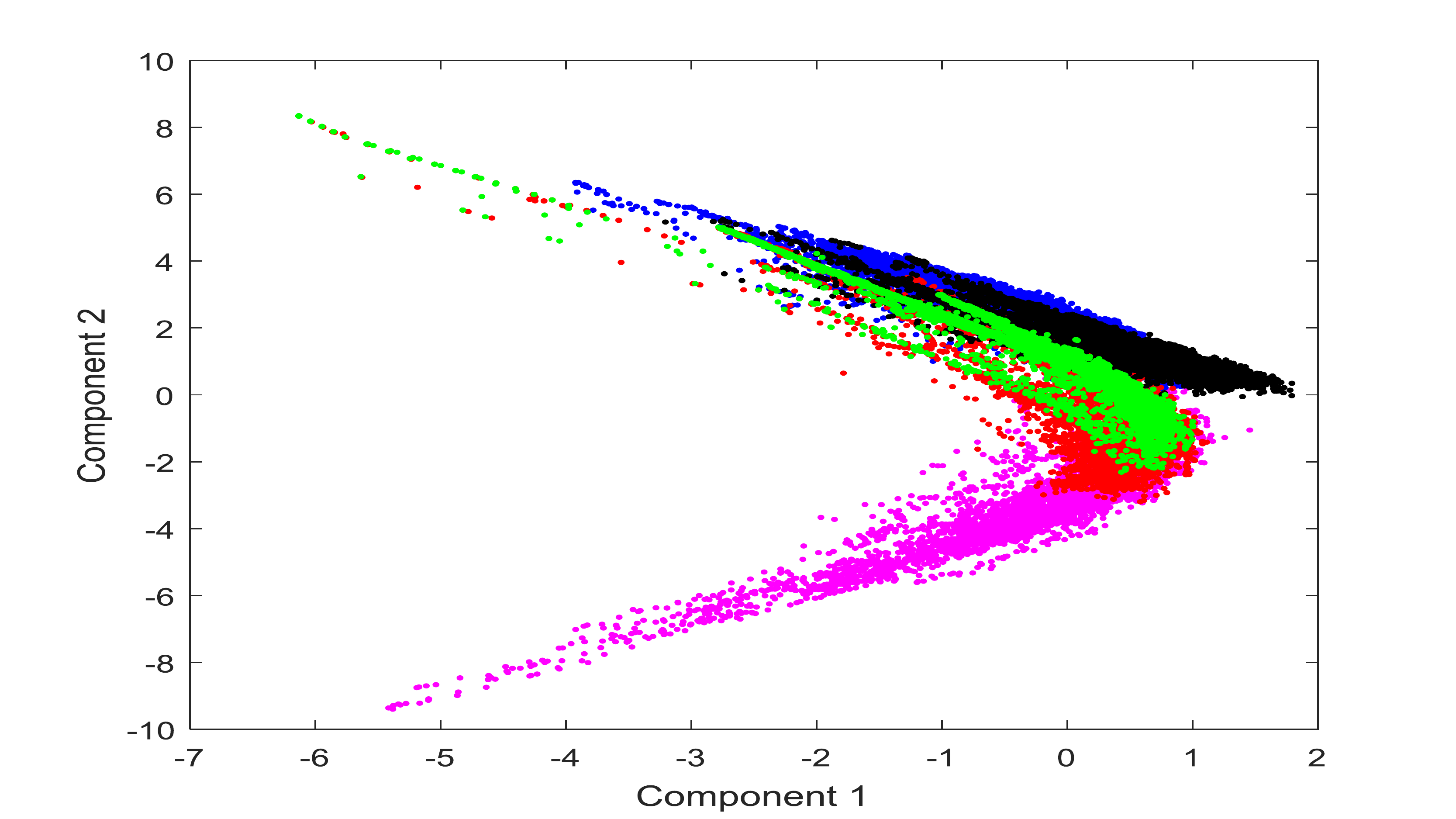}\includegraphics[width=3.3in,height=2.7in]{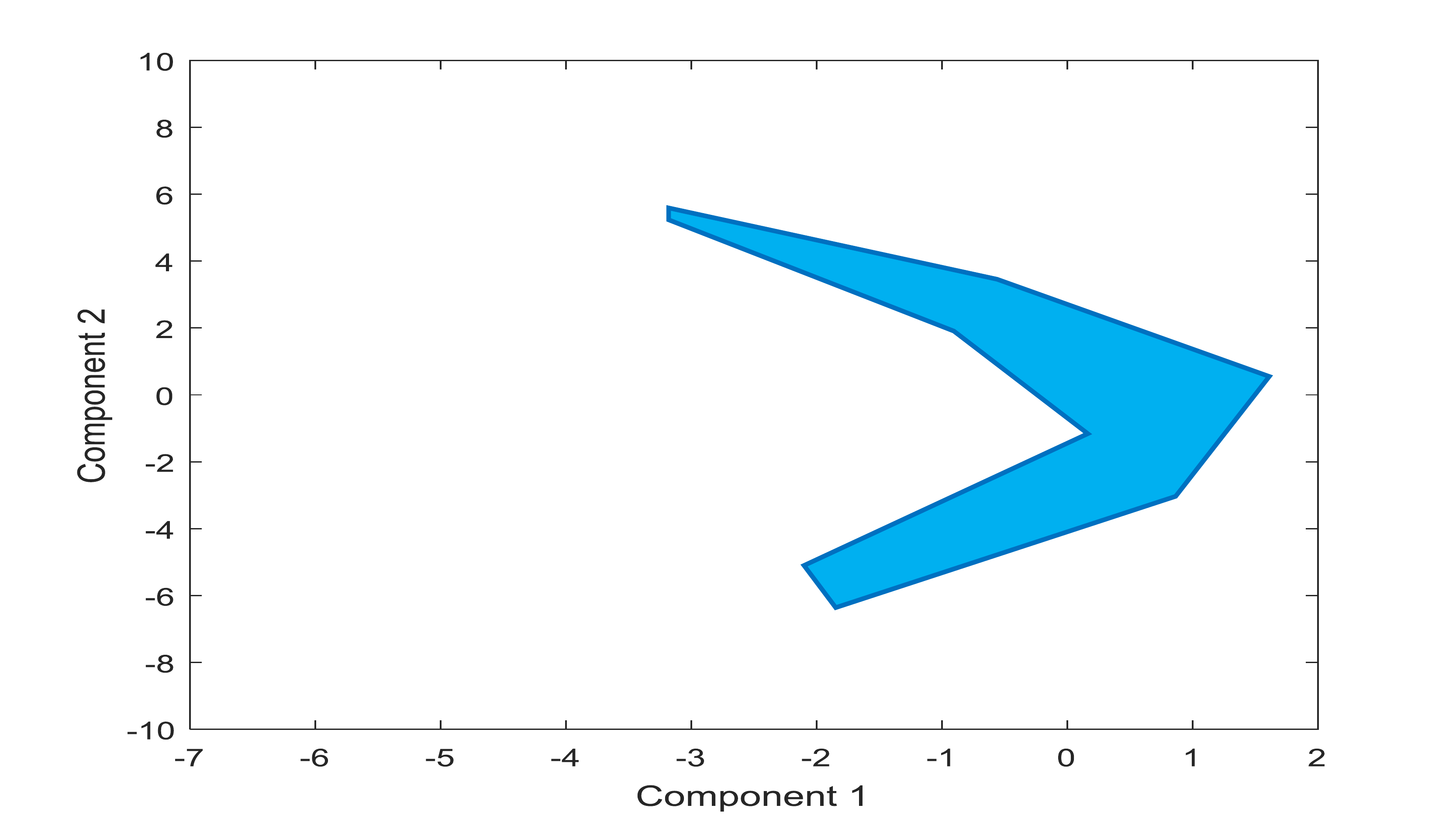}
\caption{Left: 2-d map of the RQA information computed from $6$ weeks of data, which includes the RQA variables: DET, ENT, LAM, TT, REC for the devices/device combination: table press (black), cleaner (blue), dryer (green), iron (red), aggregate (magenta). Right: The map which corresponds to the `normal' zone of energy usage at the dry cleaners.\label{clust_map} }
\end{figure*}

\begin{figure*}
\centering
\includegraphics[width=3.3in,height=2.7in]{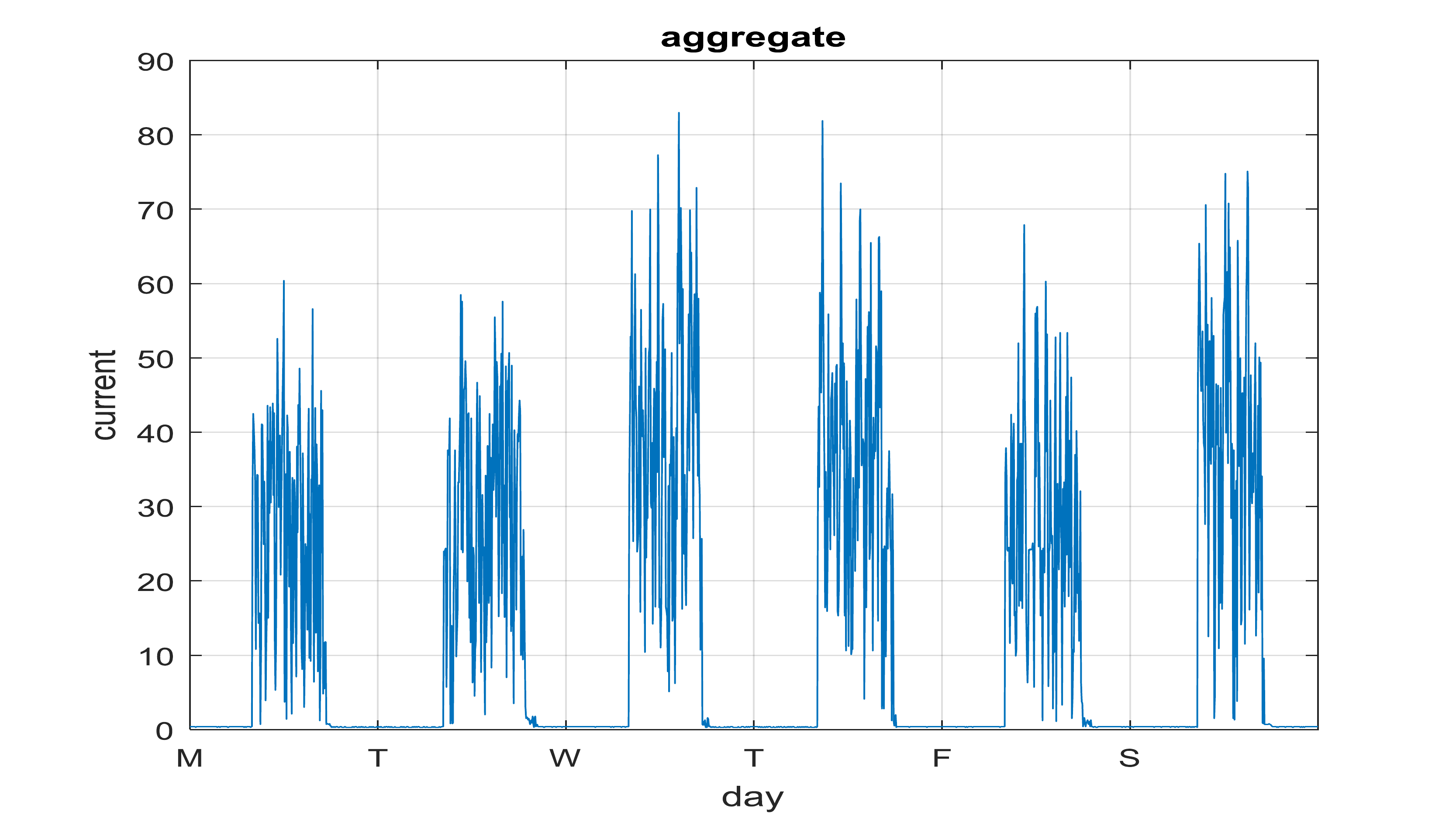}\includegraphics[width=3.3in,height=2.7in]{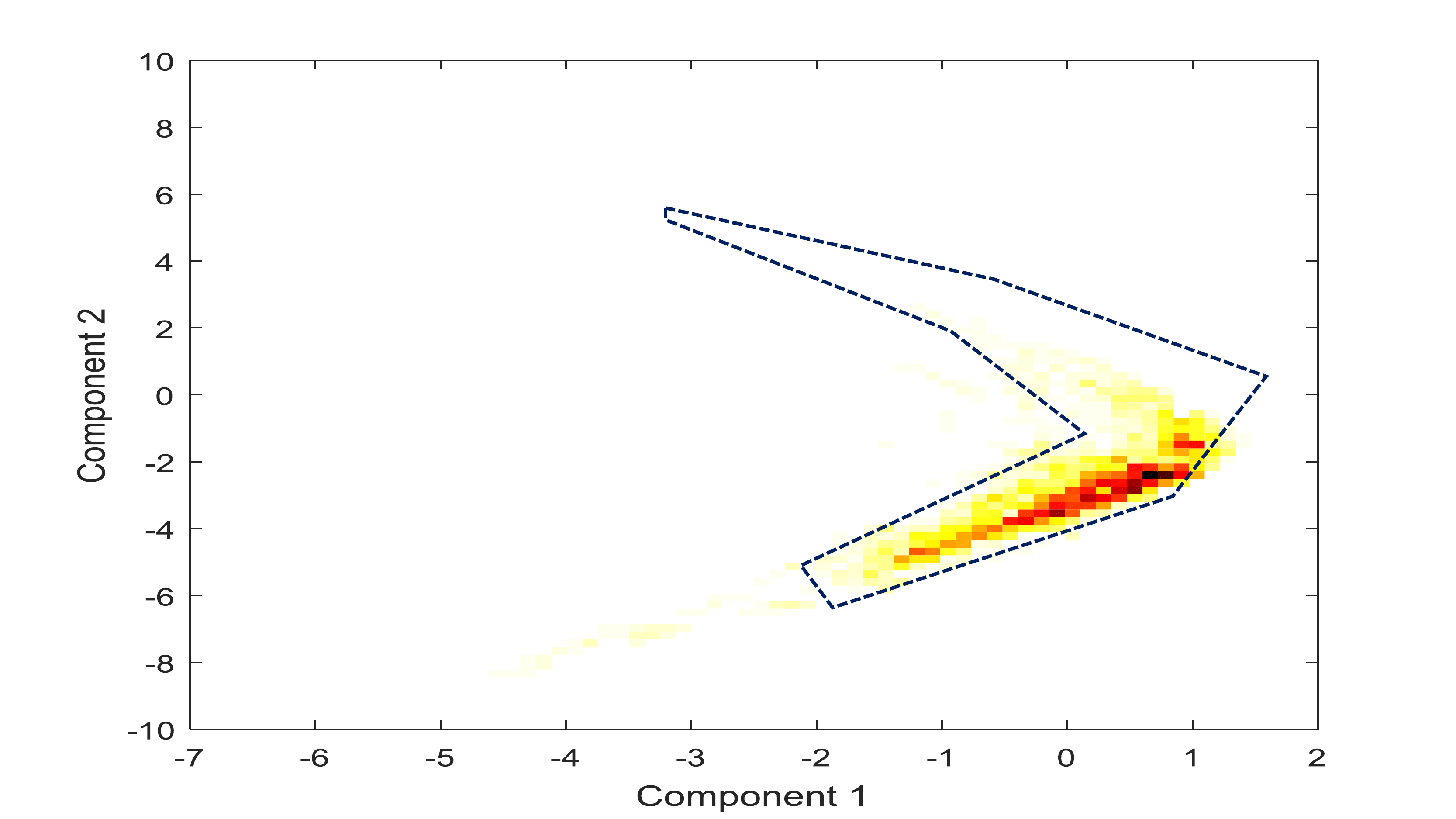}
\caption{Left: An example week of aggregate demand. Right: The signal on the left projected onto the 2-d space (darker colours correspond to increased data density), which falls outside the map $248$ times. \label{aggex} }
\end{figure*}

By applying the method outlined in Section \ref{meth}, RQA information is computed for the four dry cleaner devices and the aggregate demand, where $\epsilon=6$ Amps and $W=80$ minutes (see Sections \ref{rp}-\ref{rqa}). This information is then projected onto a two dimensional space, which is illustrated by Figure \ref{clust_map}, left. The RQA information clusters corresponding to each device and aggregate demand are highlighted in this figure, where the table press, cleaner, dryer, iron and aggregate signals correspond to the black, blue, green, red and magenta clusters respectively. Next, the information `hot-spots' for each device/device combination are determined using data density plots, which when combined form the map depicted in Figure \ref{clust_map}, right. This map signifies `normal' energy consumption at the dry cleaners and is associated with a certain energy budget. It can also be used to monitor their future demand, where newly recorded aggregate demand is projected onto this space. Then, by counting the number of data points that project beyond the map boundaries, a metric is obtained for assessing how `normal' this aggregate signal is. Moreover, if a significant portion of the data is found to fall outside the map, this can be used to infer unusual activity, which could include an additional, unexpected device switching on or a faulty appliance. Refer to Figure \ref{aggex} for an illustration of this technique, where an example week of aggregate demand (left), is projected onto the 2-d space (right) and the number of data points that fall outside the map is counted ($248$ points). Note that this signal mainly sits within the region associated with the combination of all four devices (the magenta cluster in Figure \ref{clust_map}, left). This is expected since generally at least two appliances are always switched on at the dry cleaners.

Next, so that the user can be made aware of any unexpected usage behaviour, the installation of an alarm system on site would be needed such that an alert is given when the aggregate signal starts projecting outside the map consistently. Firstly however, an appropriate threshold must be imposed corresponding to when an alarm is triggered. The method used here for calculating this threshold is explained in Section \ref{res}.

\section{Results\label{res}}
The map for monitoring energy demand at the dry cleaners has now been created (see Figure \ref{clust_map}, right). Next, this map and the proposed alarm system are applied so that the user can maintain a predetermined energy budget, where this set budget relates to daily usage in Section \ref{day} and to weekly usage in Section \ref{week}. The weekly analysis of the aggregate demand is used here to identify a faulty machine.
\subsection{A Daily Energy Budget\label{day}}
\begin{figure*}
\centering
\includegraphics[width=3.3in,height=2.7in]{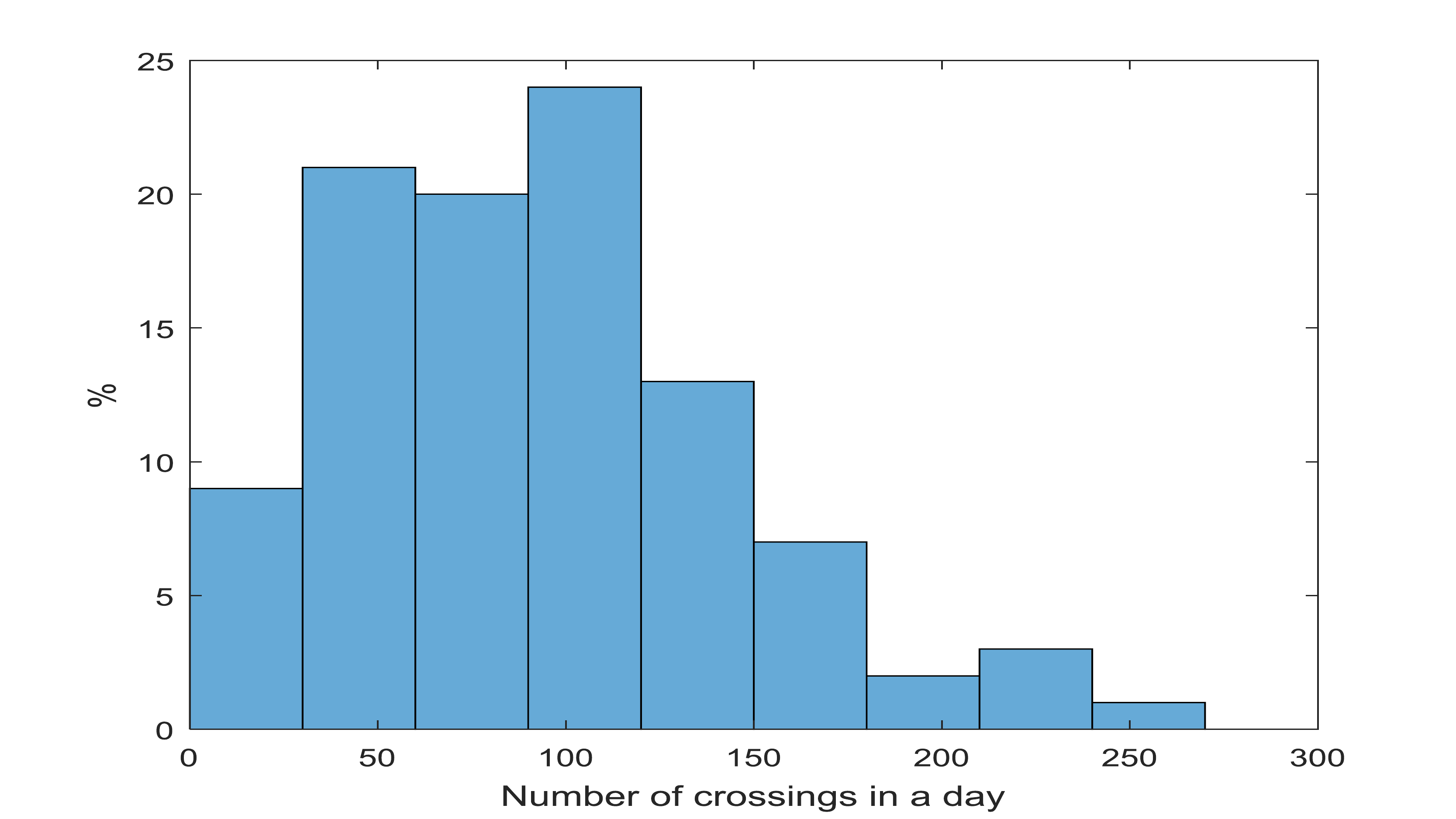}
\includegraphics[width=3.3in,height=2.7in]{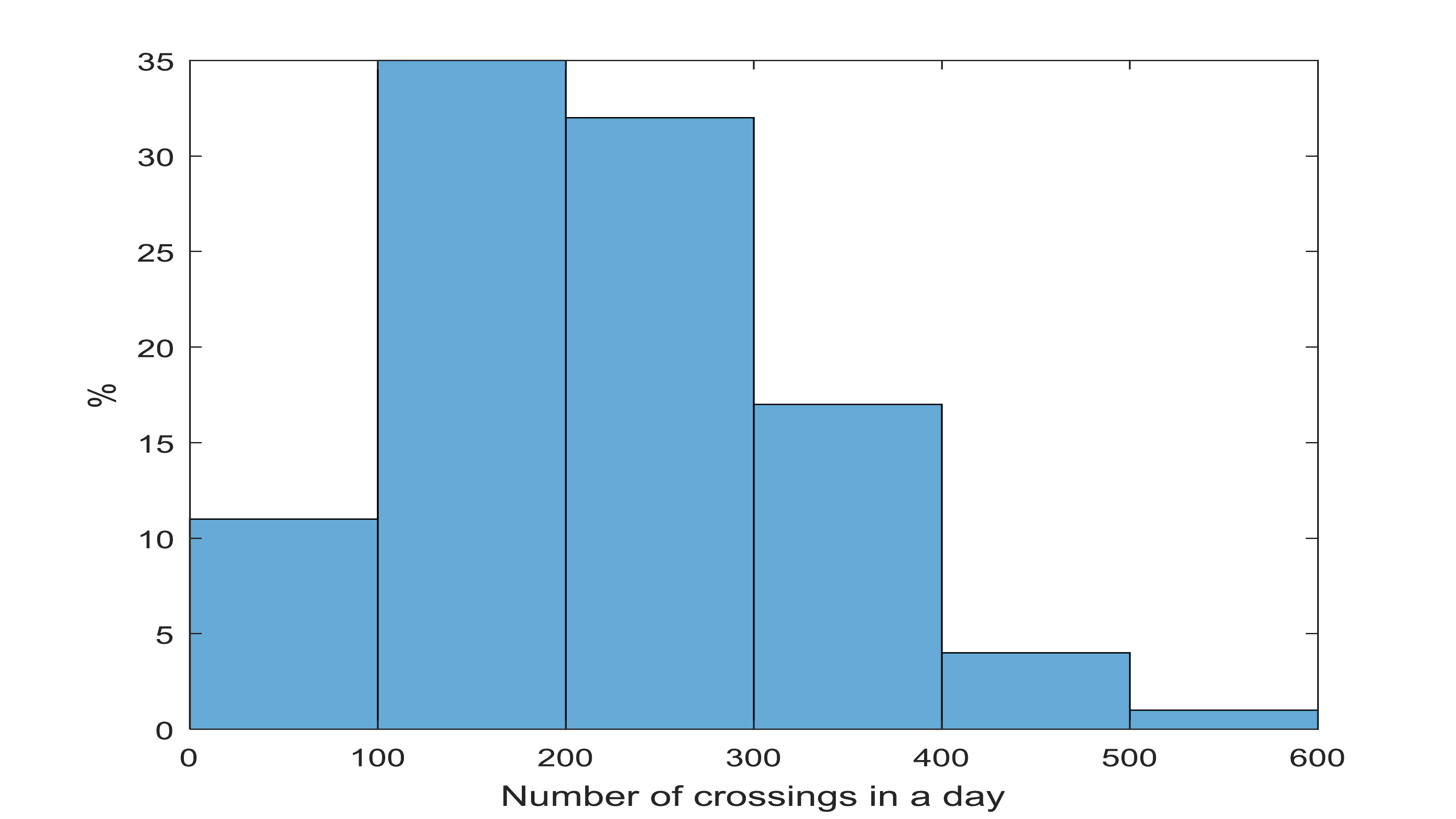}
\caption{Probability distributions for the number of daily crossings outside the map, computed from $100$ runs using the simulated daily aggregate signal: $table\; press+cleaner+dryer+iron$ (left) and $table\; press+cleaner+dryer+iron_1+iron_2$ (right). Setting the threshold to $156$ crossings outside the map in a day means that alerts are received $10\%$ (left) and $77\%$ (right) of the time.
\label{pd1}}
\end{figure*}

Firstly, to monitor and assess future daily aggregate demand, a threshold needs be set for the number of times the daily signal must project outside the map to activate an alarm. To compute this, we simulate future aggregate demand by randomly selecting daily cleaner, iron, table press and dryer profiles, summing them and then projecting this result onto the 2-d space. Next, the number of times this signal projects outside the map is counted. This is repeated $100$ times so that a probability distribution for the number of daily crossings outside our map is obtained, which is displayed in Figure \ref{pd1}, left. The threshold is then chosen as the upper bound of this distribution. Here, the alarm threshold of $156$ is chosen (the $90\%$ quantile), which corresponds to alerts being triggered $10\%$ of the time (approximately $3$ alarmed days during a $1$ month period). Hence, our map representing typical energy demand at the dry cleaners and the applied limit of $156$ daily crossings, can be used to monitor and control their daily energy consumption.

We want to now determine if this proposed method can detect the use of an additional device of comparable magnitude. The alert system can then make the dry cleaners aware that they are potentially exceeding their predetermined daily energy budget. To achieve this, multiple simulations are again performed using the steps previously outlined, except now a second randomly selected iron profile is also applied i.e. $table\; press+cleaner+dryer+iron_1+iron_2$. As a result, a probability distribution for the number of daily crossings is again found, which is depicted in Figure \ref{pd1}, right. This figure demonstrates that with an alarm threshold of $156$, the user will receive an alert $77\%$ of the time (approximately $24$ alarmed days during a $1$ month period), which suggests that the addition of another iron in the workplace is recognised and sufficient warning is received by the user. 

\subsection{Detecting a Faulty Device\label{week}}
\begin{figure*}
\centering
\includegraphics[width=2.3in,height=2.0in]{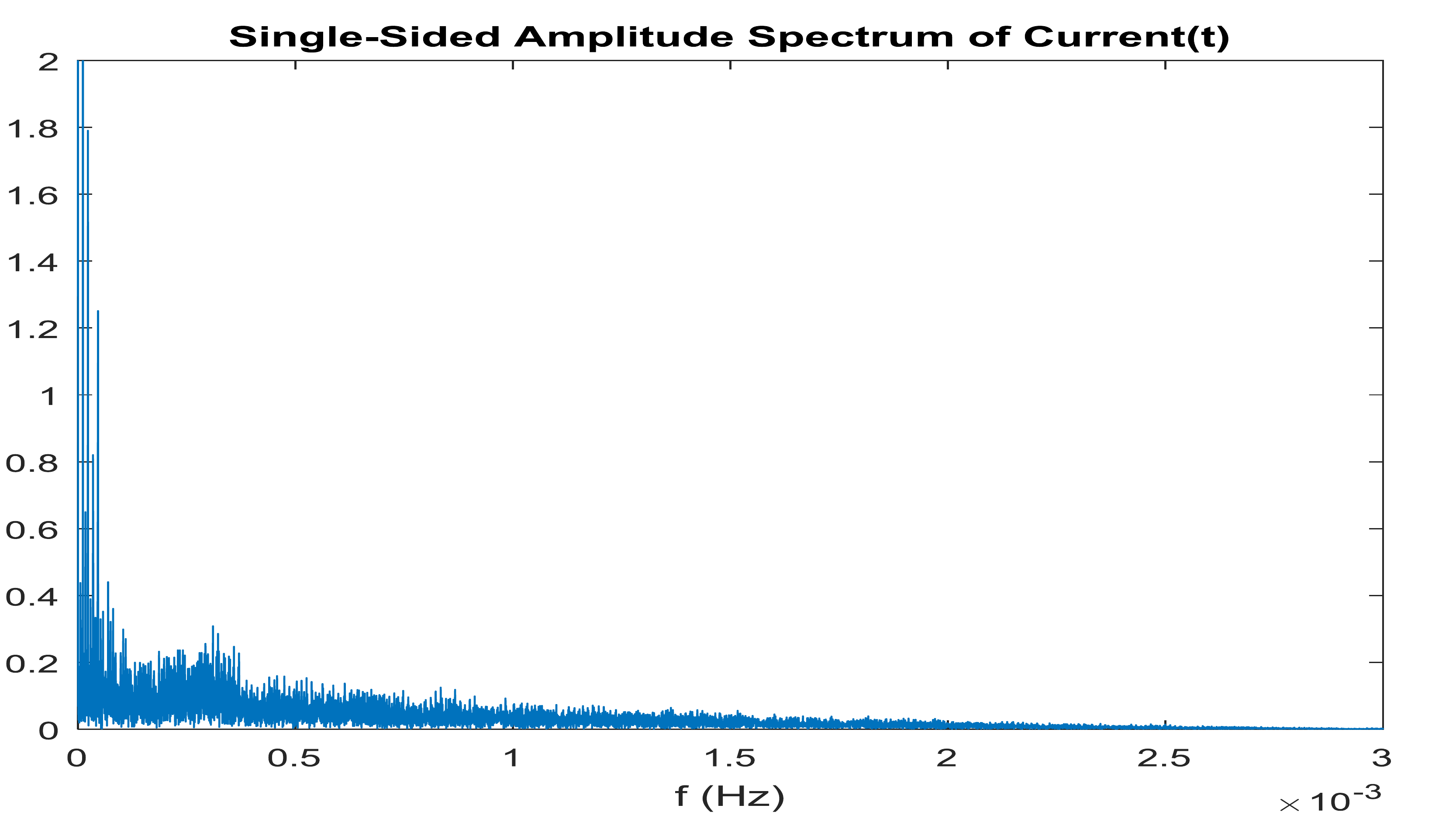}
\includegraphics[width=2.3in,height=2.0in]{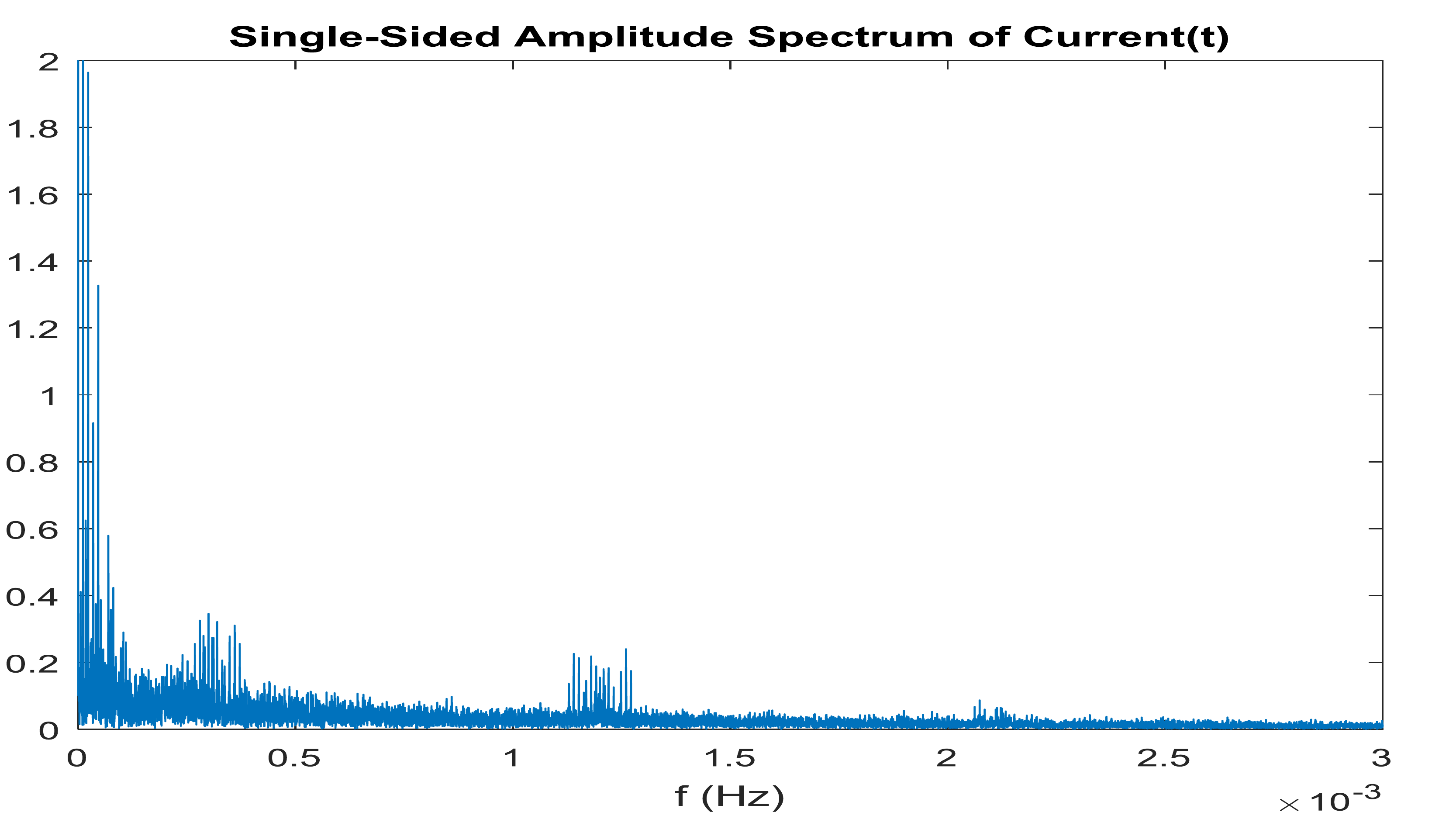}
\includegraphics[width=2.3in,height=2.0in]{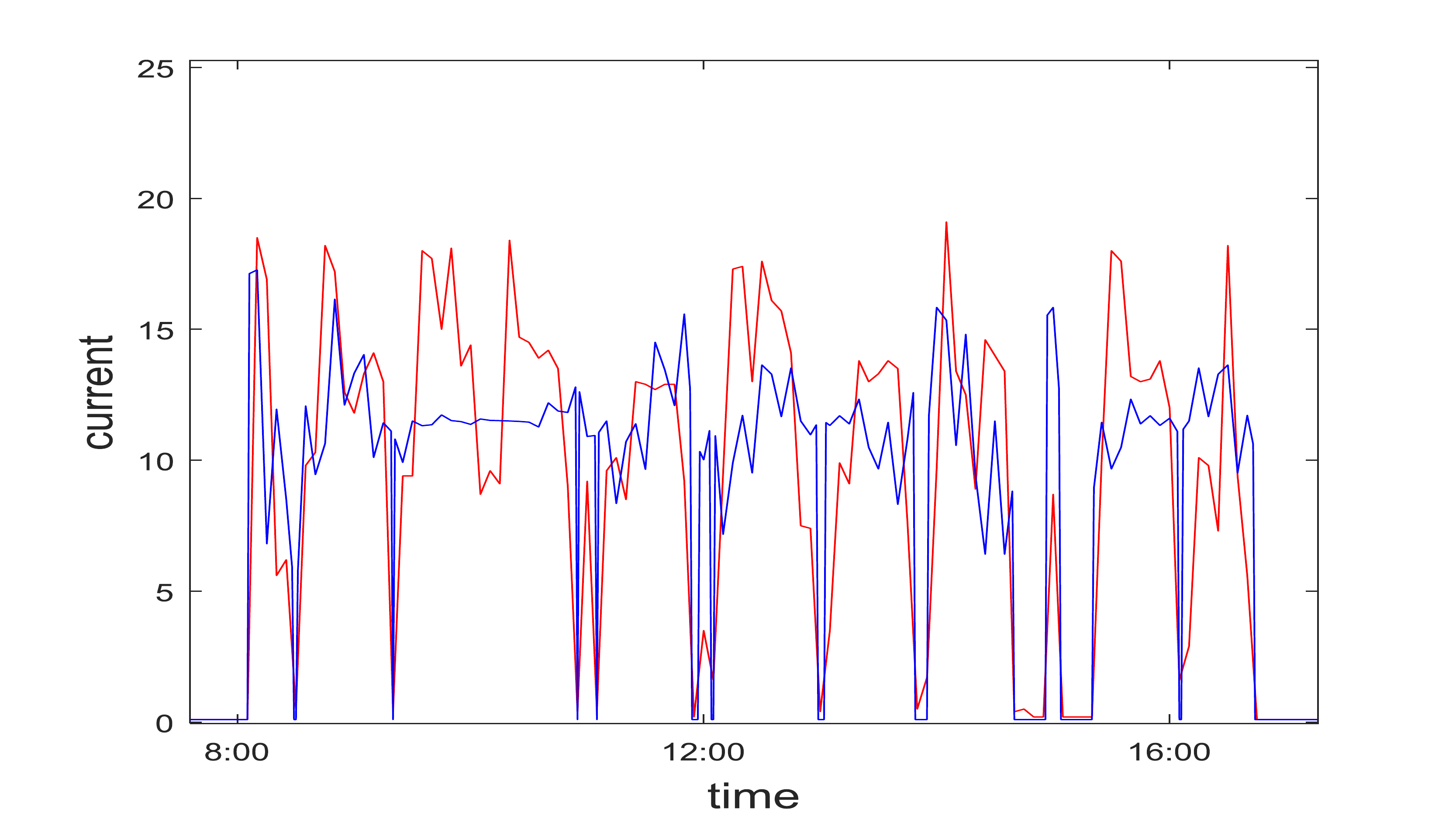}
\caption{Left and Middle: The frequency spectrum of the cleaner current signal (left) and faulty cleaner (middle). Right: Example of the daily current demands for the cleaner (red) and faulty cleaner (blue).
\label{freqs}}
\end{figure*}

Next we want to assess whether our mapping technique can detect a faulty device by simulating a cleaner with a defect. This appliance is chosen as its condition is of most concern at the dry cleaners. Numerous studies, which include \citet{tho01}, \citet{luo16}, \citet{sin17} and \citet{cul17}, have demonstrated that by analysing the frequency spectrum of a machine's current signal, various faults can be recognised due to some change in its frequency information. This approach is referred to as `motor current signature analysis'.  Therefore, here we generate a current signal for a faulty cleaner by modifying its frequency information slightly, whilst maintaining a similar amplitude, as well as consistent `on'/`off' timings. Also, this signal is initially defined every $5$ minutes, and then linearly interpolated so that it has readings every minute. The result is exhibited in Figure \ref{freqs}, where the frequency spectrum of the cleaner current signal and its faulty counterpart are depicted on the left and middle respectively. Also, Figure \ref{freqs}, right, portrays an example of the daily current demands for the different cleaner types.

\begin{figure*}
\centering
\includegraphics[width=3.3in,height=2.7in]{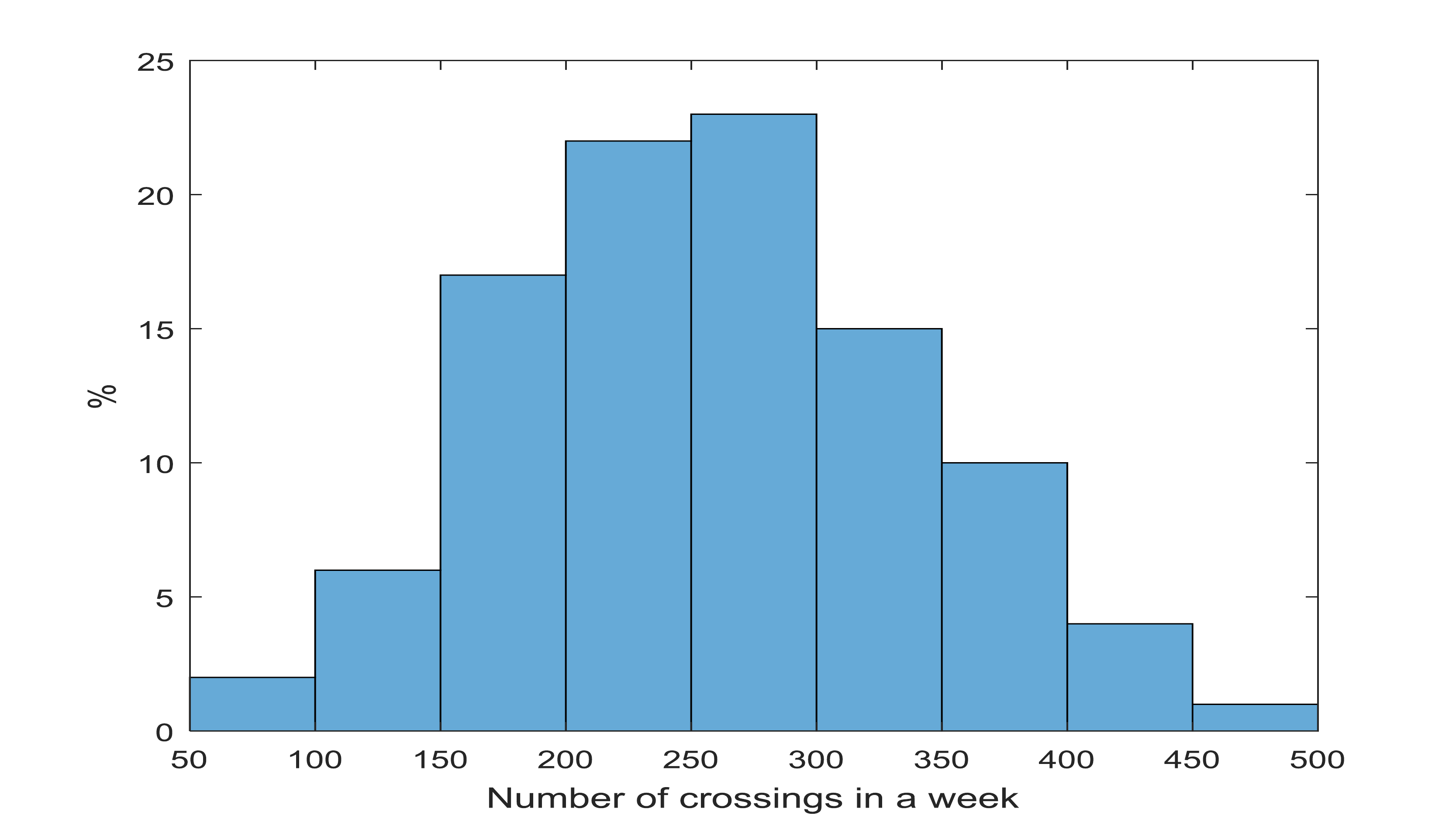}
\includegraphics[width=3.3in,height=2.7in]{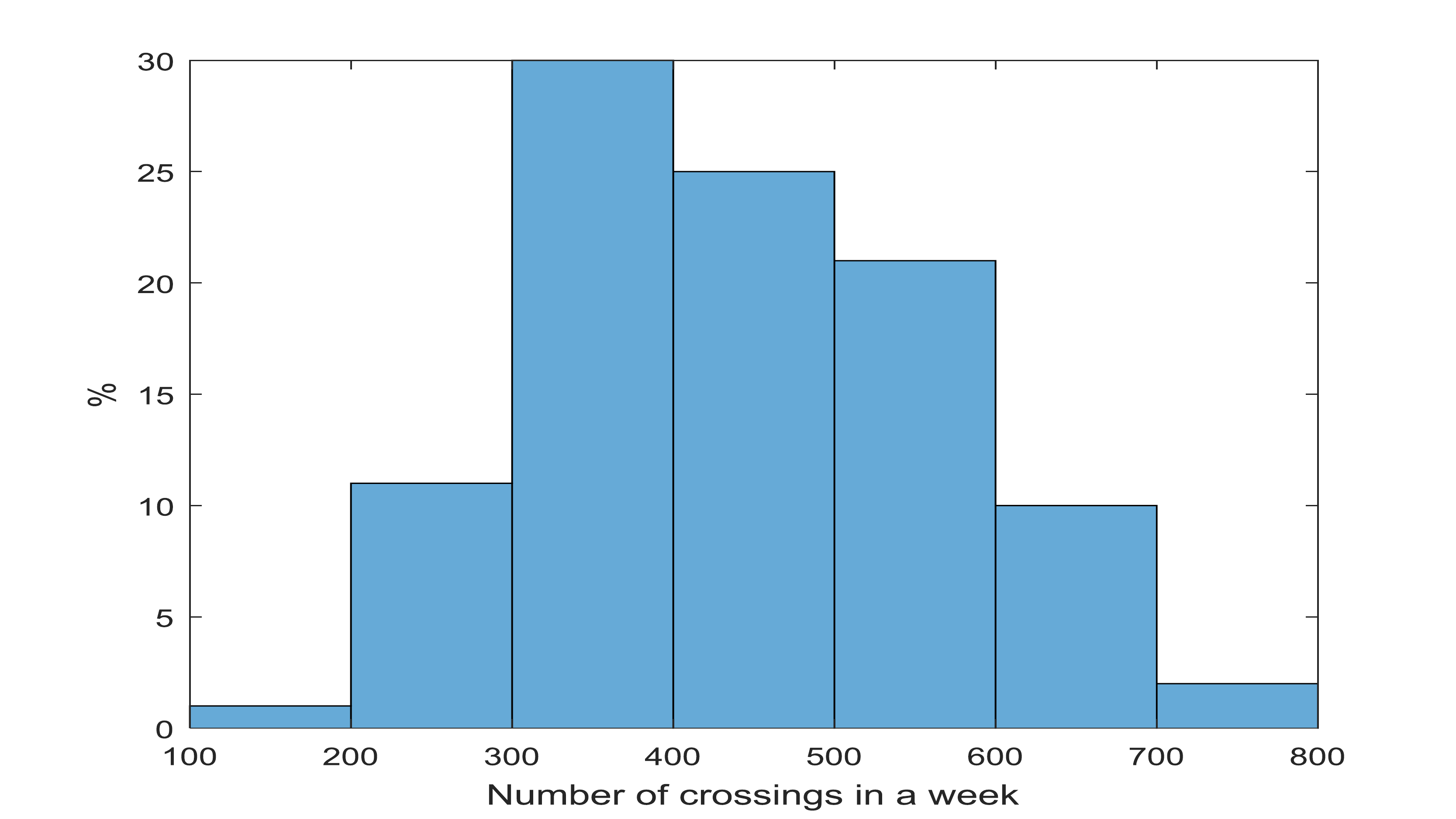}
\caption{Probability distributions for the number of weekly crossings outside the map, computed from $100$ runs using the simulated weekly aggregate signal: $table\; press+cleaner+dryer+iron$ (left) and $table\; press+faulty\;cleaner+dryer+iron$ (right). Setting the threshold to $365$ crossings outside the map in a week means that alerts are received $10\%$ (left) and $80\%$ (right) of the time.
\label{pd2}}
\end{figure*}

To use our map to now monitor machine health, a weekly alarm threshold is instead imposed. This is because the cleaner is serviced yearly, which is the only appliance that is routinely serviced at the dry cleaners, and therefore, presumably machine faults are not a daily concern here. This means that the weekly aggregate signal will be projected onto our 2-d space, and then if the number of projections outside the map exceeds the set threshold, the user is alerted. To determine this weekly threshold, the same method as previously outlined is applied so that a probability distribution for the number of weekly crossings is calculated from simulated weekly aggregate signals ($table\; press+cleaner+dryer+iron$). The resulting distribution is depicted in Figure \ref{pd2}, left. From this, the alert level of $365$ crossings is set, which corresponds to $10\%$ (approximately $5$ alarmed weeks during a $1$ year period). Next, swapping the cleaner signal with its faulty counterpart, multiple simulations are again conducted ($table\; press+faulty\;cleaner+dryer+iron$) and another distribution is computed, which is portrayed in Figure \ref{pd2}, right. This illustrates that an alert is triggered $80\%$ of the time (approximately $42$ alarmed weeks during a $1$ year period). Hence, a subtle change in one appliance's frequency information can be detected from the aggregate signal using our method. Therefore, this suggests that one appliance with an acquired defect can be identified from the total demand.

This finding will potentially have important implications for SMEs since, firstly, early detection of a faulty machine means that unexpected shut-downs are avoided, the need for routine maintenance is decreased and the operational costs are reduced. Secondly, this work suggests that the health of multiple appliances can be assessed using a single aggregate monitoring device.

\section{Conclusion\label{con}}
To conduct energy disaggregation, a dynamical systems approach was adopted, where system properties were used to differentiate between the various time series associated with different appliances. More specifically, the case study of a dry cleaners was used here to demonstrate the effectiveness of RQA for SME energy monitoring and control solutions. In particular, a user-friendly map and alert system was devised. Firstly, a 2-d map was generated that represented typical energy usage patterns at the dry cleaners, as well as corresponding to a set energy budget. Secondly, the alert system ensured that the user was made aware when unusual behaviour was detected. Next, two example applications of our proposed method were given, where the use of an unexpected device and a faulty machine triggered sufficient user warnings. Note that the machine with a defect was simulated by modifying the frequency content of its original signal. Hence, subsequent to our technique, the dry cleaners can monitor their energy usage, ensuring that they are always within a predetermined energy budget. Moreover, the health of their four main appliances can be evaluated with one aggregate monitoring device, where the fault is detected due to a change in a device's frequency information. This development has the potential to considerably reduce their operational costs, the amount of monitoring equipment and the number of routine services. Future work will include acquiring real current data from healthy and unhealthy machines so to further explore the algorithm capabilities for fault detection and its resolution limits. In addition, an automated online process will be developed to define the energy maps and the optimal alarm threshold for varying user scenarios.

\section*{Acknowledgment}
This work was carried out with the support of Innovate UK and EPSRC through the Responsive Algorithmic Enterprise project (EP/P030718/1). We thank our project partner AND Technology Research (\url{http://andtr.com/}) for providing us with the data and to Dr Valerie Lynch for sharing with us her knowledge of SME energy efficiency issues.

\bibliographystyle{plainnat2}
\bibliography{bib}




\end{document}